\documentclass[lettersize,journal]{IEEEtran}
\usepackage{amsmath,amsfonts}
\usepackage{algorithmic}
\usepackage{algorithm}
\usepackage{array}
\usepackage[caption=false,font=normalsize,labelfont=sf,textfont=sf]{subfig}
\usepackage{textcomp}
\usepackage{stfloats}
\usepackage{url}
\usepackage{verbatim}
\usepackage{graphicx}
\usepackage{cite}
\usepackage{algorithm}
\usepackage{algorithmic}
\usepackage{balance}
\usepackage{orcidlink}
\usepackage{tikz}
\usepackage{soul}
\usepackage{subfig}
\usetikzlibrary{calc}

\def \1{\textit{(i)}}
\def \2{\textit{(ii)}}
\def \3{\textit{(iii)}}
\def \4{\textit{(iv)}}
\def \5{\textit{(v)}}

\newcommand{\ie}{\textit{i.e., }}
\newcommand{\eg}{\textit{e.g., }}

\newcommand{\cf}{\textit{cf. }}

\newcommand{\sol}{PACCOR4ESP}

\begin{document}

\title{PACCOR4ESP: Embedded Device Security Attestation using Platform Attribute Certificates}

\author{\IEEEauthorblockN{
        Thomas Gr{\"u}bl\orcidlink{0009-0002-2834-3489}, 
        Jan von der Assen\orcidlink{0000-0002-0591-8887},
        Markus Knecht,
        Burkhard Stiller}

\IEEEauthorblockA{Communication Systems Group CSG, Department of Informatics IfI, University of Z{\"u}rich UZH\\
    	Binzm{\"u}hlestrasse 14, CH---8050 Z{\"u}rich, Switzerland \\ 
	E-mail: [gruebl,vonderassen,stiller]@ifi.uzh.ch}, markus.knecht2@uzh.ch
}

\author{\IEEEauthorblockN{
        Thomas Gr{\"u}bl\orcidlink{0009-0002-2834-3489}, 
        Jan von der Assen\orcidlink{0000-0002-0591-8887},
        Markus Knecht,
        Burkhard Stiller}

\thanks{This work was partially supported by (a) the University of Zürich UZH, Switzerland, and (b) the Horizon Europe Framework Program's project Certify, Grant Agreement No. 101069471, funded by the Swiss State Secretariat for Education, Research, and Innovation SERI, under Contract No. 22.00165.}
\thanks{Thomas Gr{\"u}bl, Jan von der Assen, Markus Knecht and Burkhard Stiller are with the Communication Systems Group CSG, Department of Informatics IfI, University of Z{\"u}rich UZH, Binzm{\"u}hlestrasse 14, CH---8050 Z{\"u}rich, Switzerland (e-mail: [gruebl,vonderassen,stiller]@ifi.uzh.ch; markus.knecht2@uzh.ch).}
}

\maketitle

\begin{abstract}
Verifying the integrity of embedded device characteristics is required to ensure secure operation of a device. One central challenge is to securely extract and store device-specific configurations for future verification. Existing device attestation schemes suffer from notable limitations, including a lack of standardization and a failure to encompass all hardware and software aspects inherent to a platform. This paper proposes an extension of the NSA Cybersecurity Directorate's Platform Attribute Certificate Creator (PACCOR) for the ESP32, a widely-used microcontroller series. Platform Attribute Certificates store device characteristics as per the Trusted Computing Group's Platform Certificate Profile. As of today, there is little research on hybrid attestation schemes utilizing Platform Attribute Certificates on embedded devices, which this work addresses.
  
This paper presents a collection of attacks that can be detected using PACCOR4ESP. The toolkit extracts security-relevant information from an ESP32-S3, such as the firmware hash, bootloader hash, GPIO pin configuration, and a reference to the endorsement key of the secure element, and automatically embeds it into a Platform Attribute Certificate. Lastly, this work shows how PACCOR4ESP can be integrated with existing embedded device attestation frameworks, such as RAS, CRAFT, and SEDA.
\end{abstract}

\begin{IEEEkeywords}
Device Attestation, Platform Attribute Certificates, Security, Embedded Systems, IoT
\end{IEEEkeywords}

\section{Introduction} \label{sec:introduction}
\IEEEPARstart{S}{ecurity}-critical systems often comprise multiple interconnected embedded devices. They are commonly found within Industrial Automation and Control Systems (IACS)~\cite{boyes}, the healthcare domain~\cite{alaba}, the automotive industry~\cite{bello}, and public utilities \cite{asplund}.
Due to the growing presence of such systems, the regular verification of the integrity characteristics of platforms has become a vital part of device attestation processes. Typically, the attestation process is carried out between a potentially compromised embedded device (prover) and a trusted verifier. Embedded devices are vulnerable to a range of attacks, including but not limited to firmware modification, physical, and network-based attacks \cite{Khattri, Sherman, zaddach, viega, cui, van, costin, longo}. These attacks may leave traces that can be picked up by attestation protocols.

Remote Attestation (RA) allows for verification without physical access to the device. RA techniques, commonly divided into software, hardware, and hybrid schemes, are designed to securely measure the current state of a system by remotely triggering the verification process. Software-based attestation ensures that a device's software stack remains unaltered. This may include computing checksums of the device's memory partitions. Unlike hardware-based attestation, they do not require additional hardware such as cryptographic co-processors nor physical access to the device \cite{castelluccia}. Hybrid attestation schemes combine the two aforementioned approaches and offer stronger security guarantees than software-only attestation \cite{eldefrawy2}. In cases where large-scale collective integrity verification of devices is required, so-called swarm attestation is used, a scheme that offers both security and scalability guarantees \cite{asokan}.

In recent years, numerous embedded device attestation frameworks have been proposed \cite{nunes, surminski, asokan, moreau, kim}. However, none of these consider Platform Attribute Certificates (PAC) as a means to attest to the integrity characteristics of a device. More importantly, there is no standardized approach to verifying and storing such integrity characteristics, which hinders interoperability and complicates the integration of attestation schemes. Pure software attestation schemes also suffer from security issues as they do not rely on secure hardware elements and cryptographic secrets stored within them. Co-processor-based attestation, on the other hand, is more secure since it is based on hardware secure elements. However, they are oftentimes not present in low-end embedded devices and need to be separately attached, raising cost and feasibility concerns amongst IoT solution designers \cite{ibrahim}. Yet, as \cite{pearson} point out, it is a common misconception that the security issues of Internet of Things (IoT) devices are caused by the associated cost of secure hardware, such as cryptographic co-processors. In fact, our proposed setup is not only securely designed but also affordable and thus suitable for large-scale implementations.

As of today, there is little research on the use of TCG PACs in embedded device security attestation. We aim to address this gap with the following \textbf{contributions}:
\begin{itemize}
    \item \textbf{PACCOR4ESP}~\cite{paccor4esp}:
    This paper proposes a use-case-driven approach to embedded device attestation by extending the capabilities of the NSA Cybersecurity Directorate's Platform Attribute Certificate Creator (PACCOR) \cite{paccor} for the Espressif ESP32 \cite{esp32}, a widely used microcontroller series. For our proof-of-concept implementation, we identify security-relevant characteristics of the ESP32 and demonstrate how PACCOR can be extended to integrate such characteristics into a PAC. The tool can be regarded as a single-device, hybrid attestation scheme for embedded devices (\cf Section \ref{sec:design}).
    \item \textbf{Security Considerations}: This work explores how the extended PACCOR implementation can aid in detecting embedded device-related threats. The findings reveal that PACCOR4ESP offers resistance against firmware, ELF file, GPIO pin, and non-repudiation attacks under various adversarial capabilities (\cf Section \ref{sec:security_considerations}).
    \item \textbf{Framework Integration}: We further analyze existing attestation frameworks and propose to integrate Platform Attribute Certificates into RAS \cite{kim}, CRAFT \cite{moreau} and SEDA \cite{asokan}, which can improve the integrity guarantees of a platform and aid in detecting attacks tailored to embedded devices. We specifically highlight the benefits of using PACs in existing hybrid attestation schemes for embedded devices, which has not been considered before to the best of our knowledge (\cf Section \ref{sec:framework_integration}).
\end{itemize}

First, we begin with the background (Section \ref{sec:background}) and an overview of related work in the area of embedded device attestation (Section \ref{sec:related_work}), before introducing the problem statement (Section \ref{sec:problem_statement}). Section \ref{sec:design} presents the design and implementation of PACCOR for the ESP32, which includes the verification protocol (Section \ref{sec:verification}). Section \ref{sec:runtime} presents the runtime performance evaluations. Section \ref{sec:security_considerations} discusses security considerations. Section \ref{sec:framework_integration} outlines how to integrate this new research into existing embedded device attestation frameworks, followed by a discussion of results in Section \ref{sec:discussion} and the conclusion in Section \ref{sec:conclusions}.

\section{Background} \label{sec:background}
This section discusses background information that is relevant for the subsequent sections. The main concepts required for the design of the system are Platform Attribute Certificates and the Platform Attribute Certificate Creator.

\subsection{Platform Attribute Certificates}

A Platform Attribute Certificate (PAC) is a Trusted Computing Group (TCG) defined specification. It is instantiated as an X.509 certificate with a DER-encoded ASN.1 structure and stores platform-related security properties and configurations~\cite{paccor}. First and foremost, it attests that a platform contains a unique Trusted Platform Module (TPM) and Trusted Building Block (TBB) \cite{TCG01}. A TPM is a standard that describes dedicated microcontrollers that serve as hardware trust anchors on a platform, allowing it to securely generate and store cryptographic keys.
The intended consumer of a PAC is a Privacy Certificate Authority (CA) or an enterprise that wants to verify the integrity of its devices. Typically, the intended issuer of a PAC is a platform manufacturer (\eg an Original Equipment Manufacturer (OEM)) with their own CA. Other issuing entities may be indirect device manufacturers, which act as the last trusted instances that make legitimate and authorized changes to the device configuration \cite{TCG01, TCG02}. The PAC embeds information, including but not limited to the platform manufacturer, platform model, platform version, component descriptors, and references to Endorsement Key (EK) certificates. The distinct component types recognized by a PAC are specified in the TCG Component Class Registry \cite{TCG03}. In cases where a customer wants to perform modifications to the platform themselves after it has been shipped, the PAC standard introduces the concept of delta platform certificates. Delta certificates track changes over time that could involve the addition or removal of components, adjustments to configurations, or other alterations affecting the platform's attributes. By examining the complete set of base and delta platform certificates, one can verify the history of modifications and ensure that the changes align with the trust placed in the entities performing those modifications \cite{paccor}. Both the notion of base and delta PACs help to maintain transparency and security in attestation processes.

\subsection{Platform Attribute Certificate Creator}

The Platform Attribute Certificate Creator (PACCOR) \cite{paccor} is an open-source implementation developed by the NSA Cybersecurity Directorate that supports creating and testing PACs. The implementation follows the official TCG PAC \cite{TCG01} as well as the TCG Component Class Registry \cite{TCG03} specification and is available for Windows and Unix-like systems. The tool is designed to mitigate supply chain related risks by tracking changes made to a device throughout different production stages -- until it reaches the consumer, and permanently storing these alterations in a PAC. PACCOR can be used to continuously create PACs that encapsulate the current state of a device. If the current state differs from one of the previous states, it signifies the presence of an anomaly in the device.

From an architectural perspective, PACCOR uses Powershell and Bash scripts to extract all security-relevant information from the prover and encapsulate it in the TCG-specified JSON objects. It includes Java tools that automatically generate and populate a PAC with the information stored in the JSON objects. PACCOR also requires the presence of a TPM in the system, as it will try to extract the public part of the EK, create an EK certificate, and store it as a reference in the PAC \cite{paccor}.

\section{Related Work} \label{sec:related_work}
In this section, a brief review of existing attestation methods for embedded devices is provided. Table \ref{tab:related_work} summarizes the most important attestation methods. Hybrid attestation schemes combine software-based attestation with custom hardware security extensions (\ie cryptographic co-processors) for secure hardware-based key storage. Such hardware extensions allow for unique identification of a device based on the secure element, or more specifically, the private part of the EK stored within its key storage. Hybrid schemes were first introduced to address the shortcomings of pure software-based attestation \cite{eldefrawy2}. One of the first contributions in this field was SMART~\cite{eldefrawy}, which focuses on low-end embedded devices that lack specialized memory management or protection features. It guarantees key isolation, memory safety, and atomic execution by minimally adapting the memory bus access logic in Micro-Controller Units (MCU) and establishing a dynamic root of trust.
HYDRA~\cite{eldefrawy2} is also a software/hardware co-design solution that uses the seL4 microkernel and builds upon SMART. It is designed for more powerful devices and equally guarantees memory isolation and access control. The attestation is designed on top of seL4, which offers robust security features that were traditionally achievable only through dedicated hardware.

DIALED, a Data-Flow Attestation (DFA) scheme, presented in \cite{nunes}, is specifically designed for resource-constrained embedded devices. It works hand in hand with a Control-Flow Attestation (CFA) scheme to detect various types of runtime software exploits. DIALED's primary function involves identifying and securely logging all external inputs received throughout program execution. This encompasses inputs from peripherals, the network, General-Purpose Input/Output (GPIO) pins, and data retrieved from memory locations.

RealSWATT \cite{surminski} is a continuous software-based attestation framework that performs attestation on a separate processor core on the prover to isolate the normal operation from attestation-related tasks.
RealSWATT is time-sensitive, meaning it continuously sends requests to the prover during runtime and may result in ingenuine attestation reports of the prover.

SEDA \cite{asokan} is the first attestation scheme that can efficiently attest a large swarm of low-end embedded devices and is resistant to software-only attacks. Since then, several other swarm attestation schemes have been proposed \cite{carpent, ammar, wedaj, kuang}, which improve on some of the shortcomings of SEDA.

Unlike the aforementioned frameworks, \cite{krishnan} presents a behavioral integrity verification approach by using Manufacturer Usage Description (MUD) profiles to enforce security policies and employing blockchains as the immutable storage for the IoT device network manifests. Frameworks of this type are limited to network data only and are dependent on the accuracy of the MUD profile.

One of the shortcomings of existing attestation frameworks is their implicit assumption that attackers cannot tamper with the hardware components of a platform. Additionally, some frameworks focus solely on addressing remote software-based attacks, neglecting other potential threat vectors. Another challenge is the resource-intensiveness of certain attestation processes. These limitations underscore the need for a more comprehensive and standardized approach to attestation that accounts for both hardware and software components while ensuring efficient resource utilization.

\begin{table*}[t]
\centering
  \caption{Comparison of existing attestation methods for embedded devices}
  \label{tab:related_work}
  \begin{tabular}{@{}|c|c|c|c|c|@{}}
    \hline
    \textbf{Work}&\textbf{Year}&\textbf{Category}&\textbf{Device Types}&\textbf{Special Property}\\
    \hline\hline
    SMART~\cite{eldefrawy}&2012&Hybrid attestation&Low-end embedded devices&Requires hardware modifications\\
    \hline
    SEDA~\cite{asokan}&2015&Software attestation&Low-end embedded devices&Scalable to swarms with up to 1,000,000 devices\\
    \hline
    HYDRA~\cite{eldefrawy2}&2017&Hybrid attestation&High-end embedded devices&Designed for the formally verified seL4 microkernel\\
    \hline
    DIALED~\cite{nunes}&2021&Data-flow attestation&Low-end embedded devices&Designed to detect data-only attacks\\
    \hline
    RealSWATT~\cite{surminski}&2021&Software attestation&High-end embedded devices&Designed for legacy devices under real-time constraints\\
    \hline
    PACCOR4ESP&2024&Hybrid attestation&High-end embedded devices&Leverages the TCG PAC standard\\
\hline
\end{tabular}
\end{table*}

\subsection{Problem Statement} \label{sec:problem_statement}

As touched upon in the previous sections, PACs have not yet found their way into embedded device attestation frameworks despite promising a lightweight and secure approach to device integrity. Initially designed for supply chain validation use cases, PACs are commonly shipped as part of a product between factories and customers. However, the TCG-defined PAC standard can (and should) also be used by end-users who want to attest their embedded devices continuously. Many existing device attestation frameworks \cite{carpent, darpa, salad, makhdoom} only consider a limited number of system aspects (\eg firmware-only attestation). PACs offer the opportunity to investigate a comprehensive set of device hardware and software characteristics and allow end-users to establish a trust framework for their embedded devices, ensuring ongoing security and integrity assessments.

There are frameworks that offer adequate security levels; however, since they do not comply with the TCG PAC standard, they may lack interoperability. More specifically, issues could arise when attempting to integrate attestation techniques on embedded devices that have not previously been considered. Therefore, it is beneficial to introduce standardization, a common approach to verifying device integrity characteristics across various attestation frameworks. Particularly because both the TCG PAC standard itself and PACCOR are open-source, the integrity measurement part of the protocol can be decoupled from the verification part, allowing existing frameworks to adapt PACs without requiring major modifications.

Another issue is that depending on how the verifier stores the attestation results, the results may be susceptible to tampering attacks on the verifier's side. The use of PACs makes tampering attacks less feasible because they ensure the integrity and authenticity of the attestation results, thereby upholding the overall robustness of the system against potential security threats. Although the verifier is generally considered trustworthy, storing the results in the form of a PAC has the advantage that the pre-existing attestation results remain untampered with, even if an adversary has compromised the verifier's machine. The verifier may even share the attestation results with third parties, knowing they are protected against tampering.

Furthermore, as stated in the official PACCOR \cite{paccor} documentation, PACCOR relies on additional tools to extract the underlying device data, for instance, \textit{lshw}\footnote{https://linux.die.net/man/1/lshw} for Linux to extract information on the hardware configuration. This reliance on the correctness of additional tools can be reduced when redesigning PACCOR for embedded devices, including PACCOR4ESP, where low-level functions are directly reading device configuration details. As shown in Section \ref{sec:design}, PACCOR4ESP has little impact on the underlying devices, both in terms of runtime and storage requirements.

\section{Design and Implementation} \label{sec:design}
This section focuses on the design and implementation of PACCOR4ESP using the ESP32-S3 microcontroller, a widely used embedded device for IoT applications.

PACCOR4ESP automatically generates the PAC which embeds all relevant device component details. As presented in \figurename{} \ref{fig:paccor4esp_architecture}, its main components are (1) a collection of attestation functions that extract a snapshot of the current state of the ESP32, (2) a log parser that creates the JSON files that serve as the input for the certificate creator and (3) the certificate creator itself as per the original PACCOR design, which also includes a signer and validator program. The attestation scheme of PACCOR4ESP integrates seamlessly with the original PACCOR.

\subsection{Verification Process} \label{sec:verification}

From a high-level perspective, the verification process consists of generating a new PAC and comparing it with a ground truth PAC. A PAC is generated every time after the prover supplies the verifier with the up-to-date device information details and all PACs are subsequently stored on the verifier side. The newest PAC reflects the current state of the system. In this context, ``verification'' refers to the verification of measurements taken from the prover. The verification takes place on the verifier.

All terminology used for the following description is summarized in Table \ref{tab:notations}. There are two approaches to verifying the integrity of the information stored within a PAC: First, the certificate's signature may be used as an indicator to confirm that platform characteristics remain persistent over time (referred to as $approach_1$ below). Secondly, the TCG Component Class fields in the PAC can be verified individually ($approach_2$).

To ensure the verifiability of the certificate's signature, $approach_1$ requires the use of deterministic signature schemes and access to the private key, meaning only original issuers of the PAC (\ie OEMs) may recompute and compare signatures. If the PAC signature is generated using a non-deterministic signature algorithm and/or PACs are given to a different entity for future verification, $approach_2$ should be used.

By default, the reference PACCOR \cite{paccor} implementation uses OpenSSL~\cite{openssl} with RSA-2048 to generate the signing key pair. Provided that the same signing key is used and no aspect of the platform has changed, using this approach ensures that recomputed signatures are identical to the original one for unchanged devices. If the user wishes to use a different signature algorithm, a deterministic implementation of the Digital Signature Algorithm (DSA) and ECDSA can be considered. RFC 6979~\cite{rfc6979} defines a deterministic digital signature generation procedure for DSA and ECDSA. Furthermore, it needs to be ensured that a deterministic process is used when extracting the EK certificate from the secure element.

\begin{table}[t]
  \caption{Terminology used for the Verification Process}
  \label{tab:notations}
  \begin{tabular}{@{}|c|m{5.8cm}|@{}}
    \hline
    \textbf{Symbol}&\textbf{Description}\\
    \hline\hline
    $V$&Verifier\\
    \hline
    $P$&Prover\\
    \hline
    $Cert_{PA}$&A Platform Attribute Certificate\\
    \hline
    $Cert_{EK}$&An Endorsement Key Certificate\\
     \hline
    $Cert_{GT}$&A $Cert_{PA}$ that represents the original (ground truth) system state of the prover\\
    \hline
    $SigningKey_{priv}$&The private key of the signing key pair that is generated on the verifier to sign the $Cert_{PA}$\\
    \hline
    $comp$&A component of the platform, \textit{e.g.,} the bootloader\\
    \hline
    $componentlist$&A comprehensive list of hardware and software components of the platform\\
    \hline
    $comparesig()$&A function used to compare the signatures of two $Cert_{PA}$ \\
    \hline
    $comparecomp()$&A function used to compare the values of two platform components ($comp$)\\
    \hline
    $approach_1$&Compares signatures of two $Cert_{PA}$ using $comparesig()$\\
    \hline
    $approach_2$&Individual comparison of the values of the components embedded in a $Cert_{PA}$ using $comparecomp()$\\
    \hline
\end{tabular}
\end{table}

\begin{algorithm}
\caption{Verification Protocol}\label{alg:verification}
\begin{algorithmic}[1]
    \REQUIRE $V$: trigger integrity verification request
    \REQUIRE $P$: execute integrity verification functions
    \item $P$ $\xrightarrow{\text{retrieve results}}$ $V$
    \item $V$: generate $Cert_{PA}$ and embed platform components, reference to $Cert_{EK}$
    \item $V$: sign certificate with existing $SigningKey_{priv}$
    \IF{$approach_1$}
        \STATE $comparesig(Cert_{GT}, Cert_{PA})$
    \ELSE
        \FOR{$comp$ in $componentlist$}
            \STATE $comparecomp(comp_{GT}, comp_{PA})$
        \ENDFOR
    \ENDIF
\end{algorithmic}
\end{algorithm}

The pseudocode of the \textbf{Verification Protocol} is presented in Algorithm \ref{alg:verification}. At the beginning of the attestation process, the host sends a request to the embedded device, triggering the system integrity measurement functions. The embedded device extracts all the relevant device data and returns the results back to the host. $V$ refers to the verifier, $P$ represents the prover (\textit{i.e.,} embedded system) whose integrity claims need to be verified. $Cert_{PA}$ is the PAC which embeds a reference to the $Cert_{EK}$ and, as a result, to the hardware root of trust. $SigningKey_{priv}$ represents the private key of $V$ that is used to sign the PAC. The $comparesig()$ function is used to compare the signature of the ground truth PAC $Cert_{GT}$ against the signature of a newly generated PAC $Cert_{PA}$ that represents the current state of $P$, using the aforementioned approach to verify the entirety of the contents of the PAC ($approach_1$). If the $comparesig()$ function returns true, the signature has remained unchanged, meaning that no aspect of the platform has been secretly modified.
The $comparecomp()$ function is used to compare the values of individual components ($comp$) listed in the $componentlist$. For instance, in cases where a non-deterministic signature algorithm is used. If all $comparecomp()$ function calls return true, no component of the platform has been modified.

In order to verify the authenticity of the PAC itself, the trusted certificate chain needs to be verified. This is achieved by locating the trusted issuers in the certificate chain, verifying the signatures of the certificates, and checking the validity periods of the PACs.

\subsection{Architecture and Implementation Details}

To evaluate PACCOR4ESP, we have set up a testbed consisting of the host machine \ie the verifier (running Windows \& Linux), the ESP32-S3\footnote{https://www.espressif.com/en/products/socs/esp32-s3} (\ie the prover), and the ATECC608B\footnote{https://www.microchip.com/en-us/product/ATECC608B} secure element. To communicate with the prover, we established a serial connection via the USB-to-UART bridge. We connected the ATECC608B secure element to the ESP32-S3 via the respective I2C lines.

\begin{figure*}[h]
  \centering
  \includegraphics[width=0.75\paperwidth]{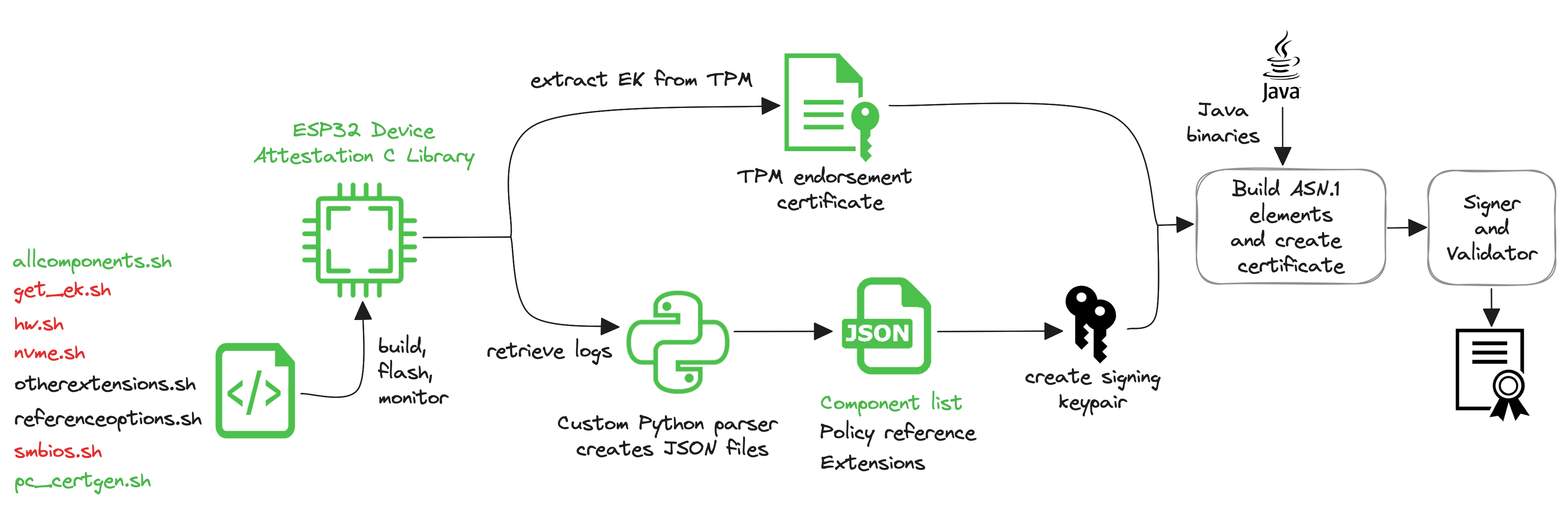}
  \caption{PACCOR4ESP Architecture, Adapted from PACCOR\cite{paccor}.}
  \label{fig:paccor4esp_architecture}
\end{figure*}

Prior to the attestation process, certain configuration variables need to be defined to ensure basic security and reproducibility of the approach. They are used by the ESP-IDF build system to customize various aspects of the program and are set in the \textit{sdkconfig} file:
\begin{itemize}
    \item \texttt{CONFIG\_SECURE\_BOOT} enables the RSA-PSS-based app and bootloader verification scheme. The signing key is a 3072-bit RSA key.
    \item \texttt{CONFIG\_APP\_REPRODUCIBLE\_BUILD} allows for achieving deterministic builds, which is imperative for comparing cryptographic hashes across different PACs. When this flag is set, build timestamps are not incorporated into the application and bootloader metadata.
    \item \texttt{CONFIG\_PARTITION\_TABLE\_CUSTOM} expects a CSV file of a custom partition table in the project's home directory. We have defined an \textit{nvs}, \textit{phy\_init} and a \textit{factory} partition.
    \item \texttt{CONFIG\_PARTITION\_TABLE\_OFFSET=0xa000} is set to accommodate for the increased bootloader size when enabling secure boot version 2.
    \item \texttt{CONFIG\_ATECC608A\_TNG} enables the Trust\&Go scheme on the ATECC608B.
    \item \texttt{CONFIG\_ATCA\_MBEDTLS\_ECDSA},\\ \texttt{CONFIG\_ATCA\_MBEDTLS\_ECDSA\_SIGN},\\ and \texttt{CONFIG\_ATCA\_MBEDTLS\_ECDSA\_VERIFY} are required to activate the Elliptic Curve Digital Signature Algorithm (ECDSA) functionality.
    \item \texttt{CONFIG\_ATCA\_I2C\_SDA\_PIN},\\ \texttt{CONFIG\_ATCA\_I2C\_SCL\_PIN},\\ and \texttt{CONFIG\_ATCA\_I2C\_ADDRESS} are set to define the Serial Data Line (SDA), Serial Clock Line (SCL), and I2C address of the ATECC608B secure element.
    \item \texttt{CONFIG\_ESP\_TLS\_USE\_SECURE\_ELEMENT} enables the secure element support.
\end{itemize}

In addition, respective \textit{sdkconfig} variables are enabled, which are required for the app itself to be functional (such as \texttt{CONFIG\_BT\_ENABLED}, which enables Bluetooth). App and bootloader rollback protection are enabled by default. \figurename~\ref{fig:paccor4esp_architecture} presents the elements and data flows of PACCOR4ESP. Elements highlighted in green represent amended or newly added components for the ESP32 use case. Black elements and data flows are part of the original PACCOR \cite{paccor} architecture. Elements highlighted in red represent components that are not relevant to the PACCOR4ESP implementation. The main entry point of PACCOR is the \textit{pc\_certgen} script, which has been adapted for the ESP32 use case. It automates the building, flashing, and monitoring process. When the program is run for the first time, it flashes the ESP32 with a sample app as well as the data extraction functions, which subsequently store the device configuration in a log file. This log file is then parsed by a Python script and transformed into the \textit{componentlist} JSON file, which serves as input for building the respective ASN.1 certificate elements. When generating a base, \ie ground truth certificate, the user can choose to generate a new signing key pair or specify an existing one in order to sign the PAC. For continuous attestation, the existing key pair should be reused, if available. In this context, continuous attestation refers to the process of regularly generating new PACs throughout the operational lifespan of a device after the initial ground truth PAC has been generated. After signing and validating the PAC, it can subsequently be decoded using an ASN.1 decoder.

\begin{table}[t]
  \caption{TCG Component Class Registry mapping to ESP32-S3 platform attributes}
  \label{tab:tcg_component_classes}
  \begin{tabular}{@{}|l|l|l|l|@{}}
    \hline
    \textbf{TCG Registry}&\textbf{Component}&\textbf{UID}&\textbf{Sample Value}\\
    \hline\hline
    0x00010008&Emb. Processor&Chip ID&2113559\\
    \hline
    0x0006000A&Flash Memory&Flash ID&295832...  \\
    \hline
    0x00090000&Ethernet Adapter&MAC&f4:12:fa:e3:91:ef\\
    \hline
    0x00090003&Wi-Fi Adapter&MAC&f4:12:fa:e3:91:ec\\
    \hline
    0x00090004&Bluetooth Adapter&MAC&f4:12:fa:e3:91:ee\\
    \hline
    0x00130003&Firmware&SHA256&c3f28aa689...\\
    \hline
    0x00130005&Bootloader&SHA256&fcc1a1e96b...\\
    \hline
    0x00130000&ELF&SHA256&93a0d81ba7...\\
    \hline
    0x00130000&Secure Boot PK&SHA256&14dd0484da...\\
    \hline
    0x000E0000&GPIO pins&Pins/Levels&1,0,0,1,0,1...\\ 
    \hline
\end{tabular}
\end{table}

The \textit{componentlist} JSON file may include information about the platform manufacturer itself, such as a manufacturer ID, platform model, version, and serial number. In addition, details about the components listed in Table \ref{tab:tcg_component_classes} are embedded into the PAC. These details may include the manufacturer name and ID, model, serial number, revision, and the component type with a reference to the official TCG registry \cite{TCG03}. Information about hardware components, such as the embedded processor, flash memory, various network adapters, and the GPIO pin states, have been extracted. We also extracted important software information, including the checksums of the firmware and bootloader partitions, the checksum of the Executable and Linkable Format (ELF) file, and the checksum of the Secure Boot V2 RSA public key. Each component has a unique identifier (UID), which is used to populate the \textit{serial} number field in the PAC. Fields such as the component revision or the component manufacturer ID are defined as optional and may be left blank.

The PAC also includes a reference to the EK certificate of the secure element. Since the ESP32-S3 does not have a dedicated secure element, such as a Trusted Platform Module (TPM), it must be extended with a hardware security module, for instance, the ATECC608B \cite{microchip} in our case. The ATECC608B is a security-enhanced version of the ATECC608A, its predecessor. It integrates the Elliptic Curve Diffie Hellman (ECDH) algorithm along with the Elliptic Curve Digital Signature Algorithm (ECDSA). The ATECC608B allows us to extract the public part of the EK, along with the signature of the private EK, and store it as a reference in the base PAC, thus irreversibly binding the EK to the ESP32, as foreseen by the TCG PAC standard.

\subsection{Generalizability}

As demonstrated through the PACCOR4ESP implementation, PACCOR \cite{paccor} is generalizable by adapting the prover software to the underlying architecture, making PACCOR a versatile attestation approach (see also Section \ref{sec:framework_integration}).
This is made possible because PACCOR decouples device-specific functionalities from the core attestation logic running on the verifier. Consequently, the approach can accommodate different prover device architectures.

While modifications are necessary to adapt the prover software to specific embedded device architectures, the fundamental PAC creation process on the verifier remains largely unchanged. On the verifier side, only the initial invocation of the attestation process and the custom log parser are required to be adapted to different embedded device platforms. The creation of the component list, policy reference, extensions JSON files, as well as the PAC creation process (as per~\figurename~\ref{fig:paccor4esp_architecture}) remain identical.

\section{Runtime Evaluations}\label{sec:runtime}

We have tested the runtime of the attestation process, which comprises both the functions executed on the ESP32-S3 to extract all device-related details as well as the functions needed to extract the EK reference from the ATECC608B secure element. We have differentiated between the two sets of functions because different hardware secure elements may yield different access times.

The runtime was measured with a microsecond precision using the internal clock of the ESP32-S3. The mean runtime of the attestation functions is averaged over 10 repetitions. Repeated runtime measurements were stopped after 10 repetitions due to negligible variability. Table \ref{tab:runtime_both} (top) presents the measured runtime performance of the attestation functions executed on the ESP32-S3, which target all device components listed in Table \ref{tab:tcg_component_classes}. In total, twelve functions are called, which log the respective results to the ESP monitor. Similarly, the runtime for extracting the EK public key information from the ATECC608B was measured (Table \ref{tab:runtime_both} bottom).

\begin{table}[t]
  \caption{Runtime performance measurements (in ms).}
  \label{tab:runtime_both}
  \begin{tabular}{|m{3cm}|l|l|l|}
     \hline
    \textbf{Measurement}&\textbf{Mean}&\textbf{Std. Dev.}&\textbf{Min/Max}\\
    \hline\hline
    Attestation functions on the ESP32-S3&395.223&0.633&394.005/395.899\\
    \hline
    EK reference extraction from the ATECC608B&148.906&0.007&148.891/148.915\\
    \hline
\end{tabular}
\end{table}

All attestation functions can run on the prover's side in under 550 ms. Unlike time-sensitive continuous attestation frameworks such as \cite{surminski}, PACCOR4ESP is not designed to run repeatedly within very short time intervals (\eg every second). Hence, the increase in the runtime of a high number of iterations will not negatively impact the intended real-world use cases. The runtime of the PAC generation process on the verifier is highly system-dependent and not crucial because the prover does not expect an immediate response. The transmission time between the prover and the verifier can be neglected.

The fully populated PAC (including all its metadata) takes up approximately 2 Kilobytes (KB) of space. Slight variations can be expected due to the contents of the PAC that may change.

\begin{figure*}[h]
    \centering
    \includegraphics[trim={0 .15cm 0 .2cm}, clip, width=.8\linewidth]{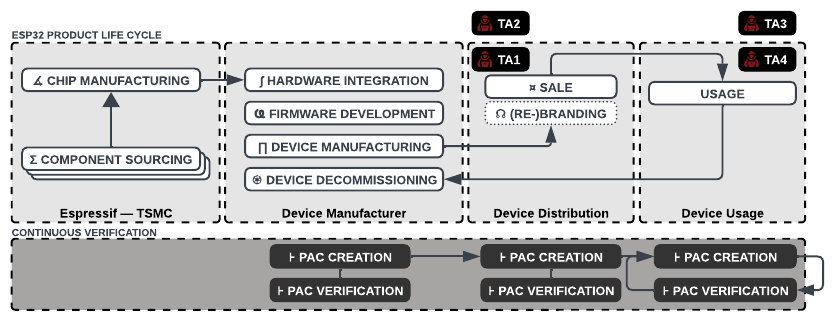}
    \caption{Illustrative Life Cycle Considered in the Security Analysis}
    \label{fig:tm-process}
\end{figure*}

\section{Security Considerations} \label{sec:security_considerations}

This section discusses the security implications of the presented approach. Since we take advantage of an already verified TCG standard, the security of this approach is not verified using formal methods. Instead, we use pre-validated components and meticulously lay out the security considerations of the overall design based on the most recent embedded device attack taxonomies. In summary, \sol{} allows for the detection of unauthorized code modifications that include firmware changes and GPIO pin configuration changes, and it ensures non-repudiation of the attestation results. It is important to emphasize that the pre-requisite of trusting the PAC is the integrity of the secure hardware element and key stored within. The following security considerations assume this pre-requisite to be true.
\subsection{System Model and Assumption}

The system model for the proposed security device operates on the assumption that the code is deployed on bare metal, ensuring the independence of specific operating systems. It is assumed that the verifier has access to a trusted high-end controller that can perform the bootstrapping process (\eg deploying the system and creating the base certificate). As a result of this process, the device is inherently known to the controller. It is essential to clarify that the security considerations within this system model pertain specifically to the communication between the provers and the verifier, excluding the security of the device bootstrapping process, which typically involves authentication and authorization mechanisms.

\begin{table}[b]
\caption{Summary of Attacks}
\begin{center}
\begin{tabular}{|ll|l|}
\hline
    & \textbf{Threat} &   \textbf{Attacker} \\
    \hline\hline
    T1 & GPIO Pin Attack      & TA3        \\
    \hline
    T2 & Device Repudiation   & TA1, TA4   \\
    \hline
    T3 & Counterfeiting       & TA2, TA3   \\
    \hline
    T4 & Firmware Tampering   & TA3, TA4   \\
    \hline
    T5 & ELF Tampering        & TA3, TA4   \\
   \hline
\end{tabular}
\label{table:summary-attacks}
\end{center}
\end{table}

\begin{figure}[b]
    \centering
    \includegraphics[width=\columnwidth, trim={0 0 5cm 0}, clip]{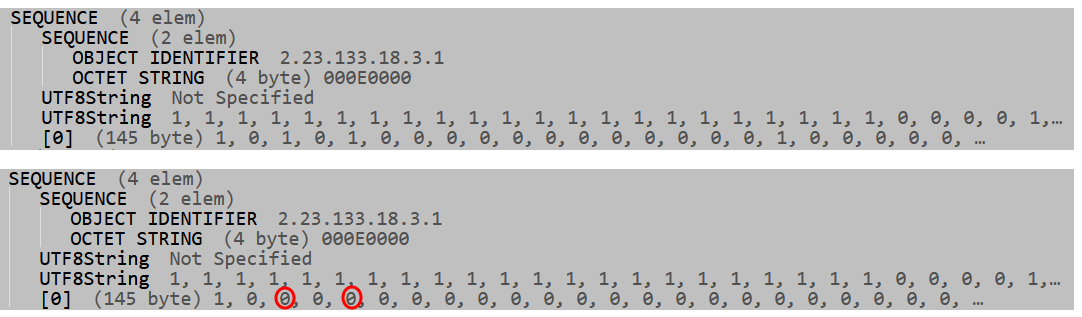}
    \caption{Attestation results after a program simulated tampering with the GPIO pin configuration and thus yielding an incorrect signature in the newly generated PAC (bottom) compared to the ground truth PAC (top).}
    \label{fig:gpio_pin_attacks}
\end{figure}

\subsection{Threat Model}

It is assumed that the verifier $V$ is a trusted entity. The prover $P$ is untrusted and vulnerable to attacks from a malicious actor $M$. Furthermore, it is assumed that $P$'s hardware root of trust (the ATECC608B secure element in our case) is not compromised.

$M$ has the following capabilities:
\begin{itemize}
    \item Can operate both locally and remotely 
    \item Can access and extensively analyze $P$
    \item Can compromise the firmware of $P$, \ie flash a compromised version into storage
    \item Can perform a physical attack on $P$, \ie attach, remove, or replace any non-secure hardware elements
\end{itemize}

Given these adversarial capabilities, a myriad of threats need to be considered. Since the work at hand aims to give the verifier $V$ insight into the state of the hardware and software of the prover $P$, the overall objective is to ensure the \textit{integrity} of $P$. Thus, the subsequent security analysis of \sol{} focuses on attacks (see Table \ref{table:summary-attacks}) that directly affect the integrity of the device or impacts that follow from such attacks (\eg data breaches impacting confidentiality after implanting a backdoor). These threats are discussed with respect to the capabilities of the approach in general.

More specifically, four adversaries are modeled using the capabilities defined above. These are contextualized using the illustrative product development process presented in \figurename~\ref{fig:tm-process}. As presented there, the first stage comprises the manufacturing process of ESP32 devices, which is trusted for the scope of this paper. The introduction of \sol{} happens in the second stage (\cf the bottom layer), where the manufacturer leverages the ESP32 chip platform to design and implement a product. Here, the vendor takes care of the integration of other hardware components (\eg sensors, actuators, or other peripherals), which must include the secure elements described in Section~\ref{sec:design}. Once the software is developed and all components are deployed, the certificate creation process of~\sol{} can be executed, after which the final product can be sold.

It is vital to reiterate that so far, the entities and components (\eg secure elements, protocols, or actors involved in the production) used by them are trusted. It is during the distribution of the product that the first two threat actors are introduced. \texttt{TA1} is either an insider attacker from the perspective of an authorized reseller or a malicious reseller who might tamper with the device's configuration. \texttt{TA2} is not a legitimate reseller but may pose as one. For example, this actor could develop a similarly looking product, which is not built on the actual hardware and software platform. In the final stage, the device is used in production with two adversaries emerging. While \texttt{TA3} is not the legitimate user of the product, he/she has access to it and tampers with the device's configuration, \texttt{TA4} is the legitimate but malicious user of the device. Nevertheless, as stated above, the verifier (\ie the current holder of the device) and the components needed for the certificate creation process (\eg ESP32 API) are considered trustworthy. As highlighted in the lower right corner of the \textit{continuous verification} layer in~\figurename~\ref{fig:tm-process}, the verifier would perform a continuous verification, which represents the main difference between the proposed ~\sol{} and the original PACCOR.

\subsection{T1: Resistance to GPIO Pin Attacks}

The GPIO pin directions and output levels can be set using the ESP-IDF API functions \texttt{gpio\_set\_direction()} and \texttt{gpio\_set\_level()} respectively. Similarly, the input level can be read using  \texttt{gpio\_get\_level()}.

Attacks similar to the one described in \cite{Khattri} can be detected by the information stored in the PAC. In the context of such attacks, it may be possible to reprogram the state of a GPIO pin, allowing an attacker to trigger unwanted alerts or set a pin to a zero value, thus blocking sensor alerts completely. To do so, an attacker (\ie TA3) would require either physical access or exploit a vulnerability of the device.
Furthermore, as described in \cite{Sherman}, GPIO pin states may be used as ``an indicator of health'', where a GPIO pin should remain in a certain state to indicate the normal operation of a system.
Since the PAC embeds information on both valid pins and the pin input levels, uncovering changes in GPIO pin configurations are possible, assuming that the attacker cannot infect the components (\ie the ESP-IDF API) on which \sol{} relies beyond re-flashing. To validate that PACCOR4ESP can detect this attack, we have modified the levels of the GPIO pins \#2 and \#4. The results are presented in~\figurename~\ref{fig:gpio_pin_attacks}.

\subsection{T2: Non-Repudiation Attack Resistance}
\label{sec:t2}
As outlined by~\cite{paccor}, supply chain security must acknowledge that multiple entities may legitimately modify components of the platform. The same observation is likely true for products built around the ESP32 chip, which could see extension or modification by different parties such as Espressif's manufacturer, the device manufacturer, the distributor, and, importantly, the consumer.

In addition to the base certificate, PACCOR can be used to create delta certificates, assuming that the intermediaries who sign them are trusted. Thus, it provides a mechanism for verifying and validating any alterations made to the configuration of an ESP32 device, effectively preventing repudiation of such changes. Throughout the manufacturing and distribution process, any entity involved, be it the manufacturer, a value-added reseller, or a distributor (\ie TA1), has the ability to introduce or remove components from the platform. Each modification made by these entities results in the generation of a delta platform certificate detailing the specific changes and authenticating them using their own certificate authority.

Furthermore, this system accommodates scenarios where the customer (\ie TA4) desires to make alterations to the platform. In such cases, the customer or a trusted office acting on his/her behalf can carry out the modifications and generate a new delta platform certificate to verify the changes. Again, the validation process relies on trusting the entity responsible for the modifications.

\subsection{T3: Counterfeiting}
Assuming that the base certificate creation is feasible to integrate into the device manufacturing process, customers and distributors can verify the authenticity of a device (\ie by verifying the EK) or certain components contained therein. More specifically, \sol{} can mitigate outsider attacks, which would forge a device and thereby deceive the customer during the device's distribution (\ie TA2) or during the decommissioning process (\ie TA4). To mitigate these risks, it is vital that the holder of the prover device is trusted. For example, if the customer wishes to verify the authenticity of the device, the customer must be a trusted actor. Similarly, if the device of a customer is to be verified by the manufacturer, it should be held by the manufacturer (unlike the scenario found in remote attestation). 

Furthermore, for these considerations, it must be assumed that the platform assessment performed by the \sol{} scripts can legitimately identify the components, revealing any potential deviation from the base certificate (if present). Therefore, attacks that can proliferate after the flashing procedure and evade detection by the inventorying component would break this security guarantee and thus are not within the scope of \sol{}.

\begin{figure}[b]
    \centering
    \includegraphics[width=\columnwidth, clip]{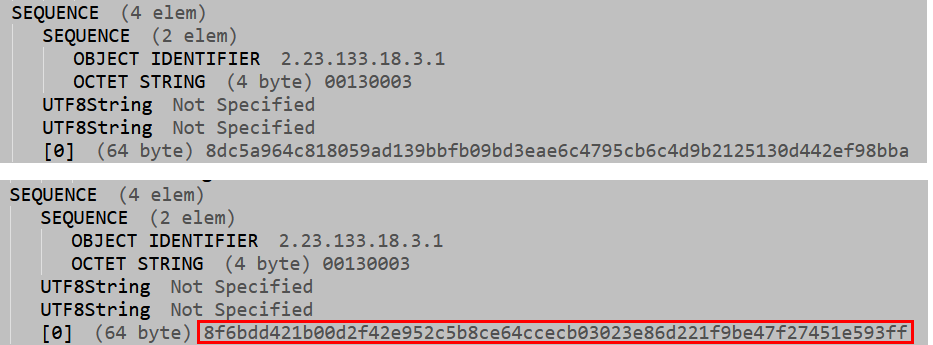}
    \caption{Attestation results after tampering with the firmware and thus yielding an incorrect firmware hash in the newly generated PAC (bottom) compared to the ground truth PAC (top).}
    \label{fig:firmware_attacks}
\end{figure}

\subsection{T4: Resistance to Firmware Tampering Attacks}
While the attack in Section~\ref{sec:t2} targeted the hardware configuration of the device, another integrity-related attack could directly attack the software flashed to the device. The differentiation between these two threat events lies in the attacker's capabilities. In both cases, the attacker would attack the device after the deployment (\ie after it was manufactured and is actually in operation). However, the attacker \texttt{TA3} or \texttt{TA4} would not necessarily require physical access to the device. Still, a local attacker, who might even be the holder of the device itself, could carry out this attack by directly modifying the software of the device. On the other hand, a remote attacker might exploit a vulnerability of the ESP32 and carry out an attack (\ie T4) that would alter the firmware. The risk exposure to this threat can be reduced by~\sol{} since the checksums of the firmware and bootloader partitions are contained in the certificates.

To validate the feasibility of detecting firmware tampering attacks, we have first generated a ground truth certificate, modified the firmware, re-flashed the device, and populated a new PAC -- capturing the current state of the device. As expected, the firmware hash stored in the new PAC differs from the one stored in the ground truth PAC (see~\figurename~\ref{fig:firmware_attacks}).

\subsection{T5: Resistance to ELF File Tampering Attacks}

\begin{figure}[b]
    \centering
    \includegraphics[width=\columnwidth, clip]{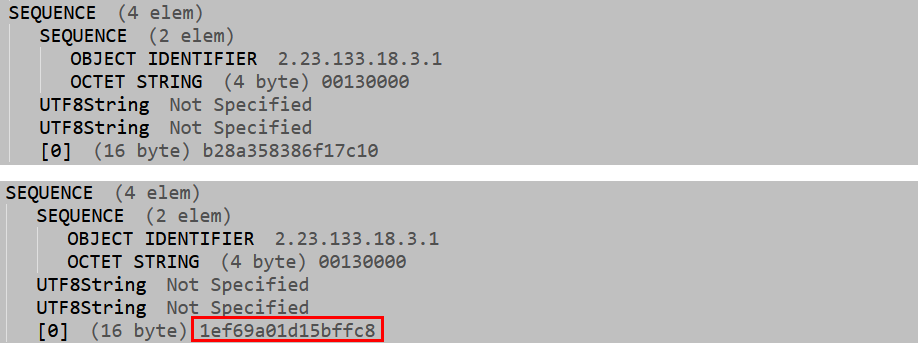}
    \caption{Attestation results after tampering with the ELF file and thus yielding an incorrect ELF hash prefix in the newly generated PAC (bottom) compared to the ground truth PAC (top).}
    \label{fig:elf_attacks}
\end{figure}

The Executable and Linkable Format (ELF) file is a standard introduced by the Tool Interface Standard Committee \cite{elf}, which specifies an object file format for Linux binaries \cite{youngdale}. There exist attacks directed at ELF files, such as ELF file injections. These attacks involve injecting code into an ELF file, causing it to load a malicious library, and subsequently restoring the regular program flow.
Similarly to T4, the PAC produced by \sol{} includes the checksum of the ELF file, which is used to build the binaries. In both cases (\ie T4 and T5), the critical assumption is that the untrusted prover can still reliably compute and report the checksums, thus stressing the necessity of the secure elements to operate \sol{}.
To simulate ELF file tampering, an additional library was included during the build process, which resulted in a change in both the ELF file as well as the firmware hash (see~\figurename~\ref{fig:elf_attacks}).

\subsection{Return-oriented Programming Attacks}

Return-oriented programming attacks are exploits that reuse existing code in the call stack of a program to craft a malicious application execution flow. This type of attack does not inject adversarial code. As described in \cite{abera}, hashes of measurements are not modified in the course of an attack that exploits control-flow vulnerabilities. Such attacks have been proven to work in many processor architectures, including Intel x86 and ARM. Since \sol{} partly relies on hash comparisons of static components, such attacks may bypass the proposed attestation approach.

\section{Framework Integration} \label{sec:framework_integration}
In this section, we analyze the feasibility of integrating PACCOR and \sol{} into existing device attestation frameworks and reason the advantages that can be gained from such an integration.
Frameworks for validating the integrity of embedded devices fall into one of two categories relevant to this work. The first category consists of frameworks where a central server validates the integrity of the devices in a network. This approach works for static systems where the central server connects to each embedded device over the network. Example use cases include systems that use a central controller to collect sensory data and issue commands to peers in the network, such as a smart device or a CCTV camera.

The second category consists of frameworks that organize devices in an interconnected swarm using peer-to-peer connections to communicate between devices \cite{asokan, carpent, ammar}. However, they do not necessarily have a reliable connection to a central server. In those networks, the embedded devices often validate each other's integrity instead of delegating this task to a central server. Nonetheless, such a network may have a central server that takes on a coordinating role and acts as a trusted entity. This approach works well for systems with mobile embedded devices, for instance, drone swarms or automated vehicles.

PACCOR is exceptionally well-suited for the first case, where the central server can use it to set up and validate the devices involved. PACCOR can only develop its full potential for the second case if the verifier can run on embedded devices. Although not originally designed to run on low-end embedded devices, PACCOR can be adapted for that context, especially when high-end embedded devices like a Raspberry Pi take on the verifier role.

Integrating PACCOR with or without PACCOR4ESP provides general benefits for frameworks from both categories and individual benefits specific to each category. The following sections discuss the general and category-specific benefits of integrating PACCOR into existing frameworks.

\subsection{PACCOR Framework Integration Benefits}
Most modern hybrid attestation frameworks rely on a cryptographic co-processor such as a TPM or a Trusted Execution Environment (TEE) in the validated device to provide a hardware-based root of trust. These frameworks use the associated certificate to ensure the integrity of the cryptographic co-processor, like the EK certificate of a TPM. However, this does not provide information on the connected device and its components. Extending these approaches with a PAC closes this gap, which is particularly useful if the devices are equipped with cryptographic co-processor extensions instead of physically integrated ones. PACCOR4ESP brings these benefits to ESP32 devices.

Using PACCOR to create a PAC for capturing properties about a device state has the added benefit that the result is TCG standardized, and even if a framework already supports a mechanism to read and store device properties, a switch to a standard such as PAC increases the interoperability with other tools, frameworks, and protocols.

Further, the PAC is a potential substitution if the targeted device does not have a suitable cryptographic co-processor. While the described attestation processes cannot provide the same guarantees as hardware-rooted solutions can, they may provide a trade-off approach to validating the device's hardware integrity. In some cases, it can be an adequate substitution, especially considering that many low-end devices do not have an integrated cryptographic co-processor. This aspect is essential, as in real-world use cases, the devices are often predetermined, and a security mechanism is selected based on the supported devices.

In the case of frameworks where a central server takes on the whole verification burden and thus acts as a root of trust, that server can use PACCOR to issue a new PAC for newly-onboarded devices during their installation, even if the OEM or other device suppliers did not issue a PAC initially. Integrating the presented validation approach using PACCOR in addition to the already present approaches can improve the integrity guarantees and aid in detecting attacks targeting embedded devices. Furthermore, verifying the signature of a PAC using the certificate of the central server allows devices in the network to identify the hardware properties of their communication partners, if necessary.

Networks based on swarm-based attestation approaches may not have a central root of trust that can use PACCOR to issue PAC certificates and, thus, have to rely on the OEM or other device supplier to issue a PAC using a public key infrastructure (PKI) to establish trust in the PAC. However, validation mechanisms can improve the security guarantees once a PAC has been generated. Especially the ability to verify the integrity of devices without a suitable cryptographic co-processor is invaluable, as swarm-based networks are often dynamic in nature and can include third-party devices that join the network temporarily; not requiring those devices to have a security-hardened hardware component can increase the number of devices capable of taking part in the system, without security concerns.

\subsection{RAS}

The Remote Attestation System (RAS) \cite{kim} belongs to the first category and requires embedded devices that use a TPM or TEE. This limitation can exclude smaller devices, such as an ESP32. Integrating PACCOR, especially PACCOR4ESP, would allow RAS to include smaller devices in their network without excluding them from the security validation process. Especially networks with many sensors where larger devices would introduce a significant overhead can benefit from this extension.

RAS has two subsystems for remote attestation (RA): the RAS RA server and the RAS RA device subsystem. The device subsystem collects information about the devices, whereas the server subsystem validates and stores them. For the integration of PACCOR, the PCA (Private Certificate Authority) component can serve as a root of trust and use PACCOR to create the PAC certificates or issue a certificate to the components that create the PACs, extending its trust to them.  

The RAS RA system has different phases that are relevant when integrating PACCOR. The system boots up in the device registration phase, where the RAS RA device subsystem can register devices with the RAS RA server subsystem. In this phase, the device subsystem can use PACCOR to create a PAC for a newly registered device and deliver it along with other information to the server subsystem. Later, the device subsystem periodically generates device status information, and at that point, PACCOR can generate a new PAC to extend that status information. In the object recovery and saving verification results phase, the server subsystem compares the information collected in the device registration phase with the periodically generated status information, which can now compare the added PACs. 

\subsection{CRAFT}

The Continuous Remote Attestation Framework for IoT (CRAFT)~\cite{moreau} is an example of a framework supporting a peer-to-peer swarm of devices. Due to its generality, it works in any network, even with a central verifier. CRAFT is designed to work with any underlying attestation protocol and can even leverage multiple protocols in the same network. CRAFT categorizes its devices into K-devices of the core network and L-devices of the outer network. The core network contains more secure devices than the outer network. The devices in the core networks can be a prover or verifier during the attestation process, whereas devices in the outer network are always in the prover role. CRAFT has two phases: an offline one, where new devices are set up, and an online phase, during which an exchange of messages happens, one of which is the attestation message that initiates an attestation process. However, it is possible to introduce additional messages if a specific attestation protocol needs them. CRAFT explicitly considers device heterogeneity and thus assumes that different devices are part of the network. CRAFT's design uses existing attestation protocols and extends them with additional functionality like a secure messaging layer and heartbeat protocols to detect malicious devices by recognizing suspicious downtime periods. 

Extending CRAFT with PACCOR and PACCOR4ESP is possible, especially due to CRAFT's focus on device heterogeneity. PACCOR4ESP would allow the integration of ESP32 devices into the CRAFT framework, when a more lightweight attestation approach is preferred. Thus, PACCOR or PACCOR4ESP may run alongside the existing protocols, increasing the range of available integrity checks. A PAC created by PACCOR can aid in the classification of the device into K and L devices by providing information about the device's hardware during the process. Devices in the core network often have more capable hardware and, in most use cases, should be capable of running PACCOR or PACCOR4ESP as a verifier or prover. ESP32 devices, on the other hand, are more likely to be in the outer network where they only need to serve as a prover, which aligns with PACCOR4ESP's intended workflow.

\subsection{SEDA}
SEDA, Scalable Embedded Device Attestation \cite{asokan} is like CRAFT, a swarm-based framework but follows a more lightweight approach. In SEDA, a central swarm operator (OP) initializes new devices, which includes issuing a digitally signed software certificate. This certificate attests to the code and software configuration that the device uses. When a device joins the network or moves to another location, it establishes a link to its neighbors, including exchanging the signed software certificate. Later, during continuous verification, a spanning tree is constructed on top of the existing neighbor relationships established during the joining phase. The tree originates from the OP or another trusted entity. Each device is a node in this tree and during verification verifies its children and forwards the result to its parent. Eventually, this information reaches the root. A single verification instance between parent and child nodes consists of the prover (child) extracting the same information stored in the software certificate and sending it to the verifier (parent), then comparing it to the signed certificate received during joining. SEDA relies on minimal hardware security requirements to ensure that the extraction of this information is not manipulatable. However, this process only attests to the running software and its configuration and does not include hardware properties.

The SEDA attestation approach between parent-child nodes has many parallels to PACCOR; in both cases, information is extracted and packed into a certificate, which then is compared to an earlier one signed by a trusted entity. The main difference is that a PAC also contains hardware information, whereas SEDA's certificate only contains software information. Thus, an integration of PACCOR is possible by reusing SEDA's existing processes. First, during device initialization, the OP can use PACCOR to create a signed PAC and store it alongside the software certificate. During joining, the devices exchange the PAC and the software certificate. Finally, PACCOR extracts and compares a new PAC to the signed one during attestation. This approach is considered secure, as the software certificate already proves that the correct software runs.

Despite those parallels, integrating ESP32 devices and PACCOR4ESP into a SEDA network can be challenging, as it requires the ESP32 to be able to run the prover and the verifier software, which is not yet the case; currently, only the prover can run on an ESP32. However, even in this case, having a PAC issued by the OP available on each device providing hardware information about its neighbor's initial state may be valuable even if not continuously verified.

\subsection{Integration Remarks}
The previous two sections show how PACCOR and PACCOR4ESP can extend existing frameworks to increase their capabilities. RAS and CRAFT separate the attestation of a concrete device from the overall approach to ensure the integrity of a system. However, not all frameworks follow this approach and some may require the attestation process to fulfill specific criteria, such as that it can run inside a TEE, use a specific code format, or must have specific performance characteristics. RealSWATT \cite{surminski} (see Section \ref{sec:related_work}) is an example of an approach that relies on a performance-optimal prover implementation that cannot be significantly sped up by parallelization or optimizations. Integration would require major modifications to PACCOR4ESP and further analysis of the results before the integration can be considered secure. Another example is DIALED \cite{nunes}, which instruments the assembler code of the executed program; thus, interpreted code, like bash scripts or Python, does not work with DIALED. However, even in cases where PACCOR is not a viable integration into the integrity attestation process of a framework or system, it is recommended to use it during device setup to generate a standardized PAC to capture device properties, especially if they are in use by the attestation protocol.

\section{Discussion} \label{sec:discussion}

This section discusses viewpoints beyond the technicalities of our approach. In the context of hybrid embedded device attestation frameworks, interoperability is essential to ensure that devices from different manufacturers can communicate and attest to their integrity in a consistent manner. \cite{bicaku} highlight the significance of adhering to standards and continuous attestation in IoT systems. This is particularly important in heterogeneous environments where devices with different architectures and functionalities need to co-exist. Moreover, standardization facilitates the scalability of IoT systems because it simplifies the management and integration of devices. The TCG PAC standard is a suitable standard for attestation frameworks for several reasons: (1) It provides consistency, renders attestation results into a well-known representation (X.509 certificates), (2) it provides a list of comprehensive attributes for all possible software and hardware components that can appear in a platform, and  (3) the TCG itself is a consortium that has gained broad acceptance \cite{TCG04}. The adoption of a TCG standard enhances the likelihood of widespread acceptance and interoperability of an attestation approach.

From a cost perspective, the proposed hybrid attestation approach is even feasible for legacy embedded devices without a pre-existing hardware root of trust, as they can be extended with an external secure element, such as the ATECC608B, which is remarkably affordable with a price of less than 1 USD\footnote{\url{https://www.microchip.com/en-us/product/ATECC608B}}. It becomes a cost-effective strategy to enhance the security of embedded devices with hardware secure elements, TCG PACs, and a customized version of PACCOR, both of which are open-source, as previously mentioned.

\textbf{Limitations:} At present, PACCOR4ESP is tailored to the ESP32 platform, relying heavily on ESP-IDF API functions. Nevertheless, the security considerations and the proposed framework integrations are transferable to other embedded devices. To generalize the approach to other platforms, the attestation software running on the prover needs to be adjusted accordingly. It is also worth noting that PACs have inherent limitations; for instance, they are unable to identify additional hardware components that might be introduced into the system, such as dedicated spyware microchips, because the attestation functions can only analyze known platform components. Furthermore, it is essential to note that PACCOR4ESP may inherit certain constraints from its predecessor, PACCOR.

\section{Conclusions} \label{sec:conclusions}

In this article, we have presented PACCOR4ESP, a single-device hybrid attestation scheme that leverages TCG Platform Attribute Certificates and PACCOR as a means of attesting to the integrity of an embedded device, including all its components and peripherals. We demonstrated the feasibility of using PACs for embedded device attestation with our proposed approach, which is designed to execute the prover software on ESP32 microcontrollers. It is not only securely designed, but also affordable and thus suitable for large-scale implementations. In the context of security considerations, we analyzed possible security threats such as GPIO pin attacks, device repudiation, and firmware tampering and showed that such malicious modifications could be picked up by the attestation process. We further analyzed various embedded device attestation schemes and proposed to integrate PAC-based attestation into three existing frameworks, the Remote Attestation System (RAS), the Continuous Remote Attestation Framework for IoT (CRAFT), and Scalable Embedded Device Attestation (SEDA), to enhance the overall versatility of the approach. It can be concluded that TCG PAC-based attestation provides a secure and standardized way of storing attestation results for embedded devices, while consuming minimal resources during the attestation process. Future work may include customizing PACCOR for other embedded device platforms.

\section*{Acknowledgments}
This work was partially supported by (a) the University of Zürich UZH, Switzerland, and (b) the Horizon Europe Framework Program's project Certify, Grant Agreement No. 101069471, funded by the Swiss State Secretariat for Education, Research, and Innovation SERI, under Contract No. 22.00165.

\bibliographystyle{IEEEtran}
\bibliography{base}

\balance

\end{document}